\documentstyle[aps,prl,epsf,floats]{revtex}
\begin{document}
\draft
\title{COMPOSITE FERMIONS AND THE \\ FRACTIONAL QUANTUM HALL EFFECT}
\author{
   \sc
   A. W\'ojs$^{ab}$ and J. J. Quinn$^a$}
\address{
   \rm
   $^a$Department of Physics, University of Tennessee, 
   Knoxville, Tennessee 37996, USA\\
   $^b$Institute of Physics, Wroclaw University of Technology,
   50-370 Wroclaw, Poland}
\maketitle
\begin{abstract}
The mean field (MF) composite Fermion (CF) picture successfully predicts 
low lying states of fractional quantum Hall systems.
This success cannot be attributed to a cancellation between Coulomb 
and Chern--Simons interactions beyond the mean field and solely depends
on the short range (SR) of the Coulomb pseudopotential in the lowest 
Landau level (LL).
The class of pseudopotentials for which the MFCF picture can be applied
is defined.
The success or failure of the MFCF picture in various systems 
(electrons in excited LL's, Laughlin quasiparticles, charged 
magneto-excitons) is explained.
\end{abstract}

\section{Introduction}

The quantum Hall effect (QHE)\cite{klitzing,tsui}, i.e. the quantization 
of Hall conductance of a two-dimensional electron gas (2DEG) in high 
magnetic fields at certain filling factors $\nu$ ($\nu^{-1}$ is the number 
of single particle states in a LL per electron), signals the appearance 
of (incompressible) nondegenerate ground states (GS's) separated from the 
continuum of excited states by a finite gap.
At integer $\nu=1$, 2 \dots\ (IQHE), the excitation gap is the single 
particle cyclotron gap, while at fractional $\nu=1/3$, 1/5, 2/5 \dots\ 
(FQHE) electrons partially fill a degenerate (lowest) LL and the formation 
of incompressible GS's is a many body phenomenon revealing the unique 
properties of Coulomb interaction of electrons in the lowest LL
\cite{laughlin,prange}.

In the mean field (MF) composite Fermion (CF) picture\cite{jain,lopez},
in a 2DEG of density $n$ at a strong magnetic field $B$, each electron 
binds an even number $2p$ of magnetic flux quanta $\phi_0=hc/e$ (in form 
of an infinitely thin flux tube) forming a CF.
Because of the Pauli exclusion principle, the magnetic field confined
into a flux tube within one CF has no effect on the motion of other CF's, 
and the average effective magnetic field $B^*$ seen by CF's is reduced, 
$B^*=B-2p\phi_0n$.
Because $B^*\nu^*=B\nu=n\phi_0$, the relation between the electron and 
CF filling factors is
\begin{equation}
   (\nu^*)^{-1}=\nu^{-1}-2p.
\end{equation}
Since the low band of energy levels of the original (interacting) 2DEG 
has similar structure to that of the noninteracting CF's in a uniform 
effective field $B^*$, it was proposed\cite{jain} that the Coulomb
charge-charge and Chern--Simons (CS) charge-flux interactions beyond 
the MF largely cancel one another, and the original strongly interacting 
system of electrons is converted into one of weakly interacting CF's.
Consequently, the FQHE of electrons was interpreted as the IQHE of CF's.

Although the MFCF picture correctly predicts the structure of low energy
spectra of FQH systems, the energy scale it uses (the CF cyclotron energy
$\hbar\omega_c^*$) is totally irrelevant.
Moreover, since the characteristic energies of CS ($\hbar\omega_c^*
\propto B$) and Coulomb ($e^2/\lambda\propto\sqrt{B}$, where $\lambda$ 
is the magnetic length) interactions between fluctuations beyond MF scale
differently with the magnetic field, the reason for its success cannot 
be found in originally suggested cancellation between those interactions.
Since the MFCF picture is commonly used to interpret various numerical 
and experimental results, it is very important to understand why and
under what conditions it is correct.

In this paper, we use the pseudopotential formalism\cite{haldane1,wojs1} 
to study the FQH systems.
It is shown that the form of the pseudopotential $V(L')$ [pair energy vs. 
pair angular momentum] rather than of the interaction potential $V(r)$, 
is responsible for the incompressibility of FQH states.
The idea of fractional parentage\cite{shalit} is used to characterize 
many body states by the ability of electrons to avoid pair states with 
largest repulsion.
The condition on the form of $V(L')$ necessary for the occurrence of FQH 
states is given, which defines the class of SR pseudopotentials to which 
MFCF picture can be applied.
As an example, we explain the success or failure of MFCF predictions 
for the systems of electrons in the lowest and excited LL's, Laughlin 
quasiparticles (QP's) in the hierarchy picture of FQH states
\cite{haldane2,sitko1}, and charged excitons in a 2D electron-hole 
plasma\cite{wojs2}.

\section{Numerical Studies on the Haldane Sphere}

Because of the LL degeneracy, the electron-electron interaction in the 
FQH states cannot be treated perturbatively, and the exact (numerical) 
diagonalization techniques have been commonly used.
In order to model an infinite 2DEG by a finite (small) system that can 
be handled numerically, it is very convenient to confine $N$ electrons
to a surface of a (Haldane) sphere of radius $R$, with the normal magnetic 
field $B$ produced by a magnetic monopole of integer strength $2S$ (total 
flux of $4\pi BR^2=2S\phi_0$) in the center\cite{haldane2}.
The obvious advantages of such geometry is the absence of an edge and 
preserving full 2D symmetry of a 2DEG (good quantum numbers are the total 
angular momentum $L$ and its projection $M$).
The numerical experiments in this geometry have shown that even relatively 
small systems that can be solved exactly on a small computer behave in 
many ways like an infinite 2DEG, and a number of parameters of a 2DEG 
(e.g. characteristic excitation energies) can be obtained from such small 
scale calculations.

The single particle states on a Haldane sphere (monopole harmonics) are 
labeled by angular momentum $l$ and its projection $m$\cite{wu}.
The energies, $\varepsilon_l=\hbar\omega_c[l(l+1)-S^2]/2S$, fall into
degenerate shells and the $n$th shell ($n=l-|S|=0$, 1, \dots) corresponds 
to the $n$th LL.
For the FQH states at filling factor $\nu<1$, only the lowest, spin 
polarized LL need be considered.

\begin{figure}[t]
\epsfxsize=4in
\centerline{\epsffile{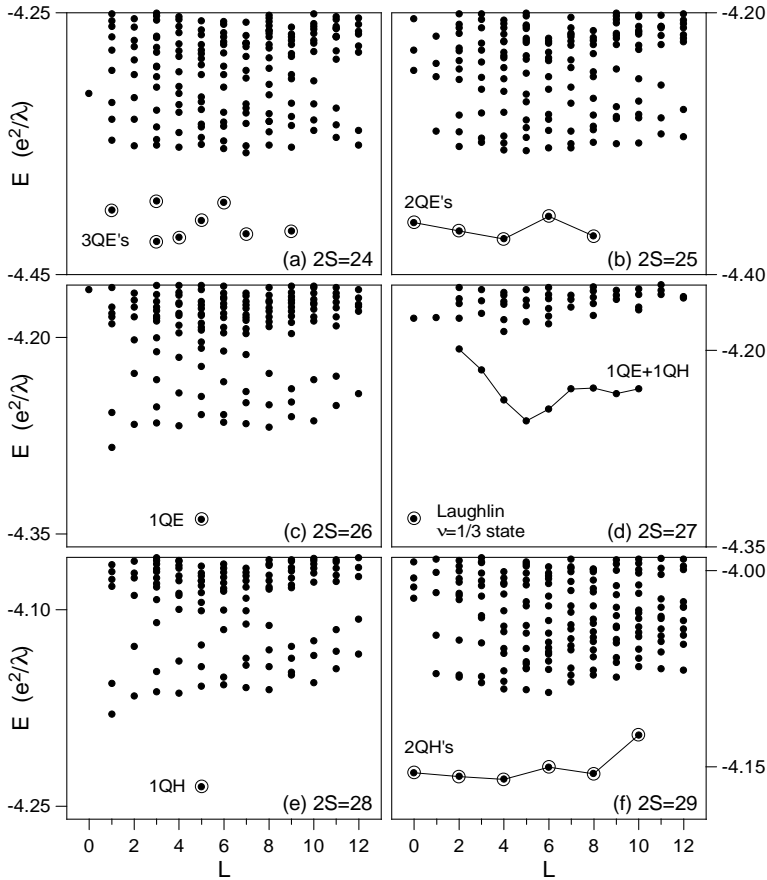}}
\caption{
   Energy spectra of ten electrons in the lowest LL at the monopole 
   strength $2S$ between 24 and 29. 
   Open circles mark lowest energy bands with fewest CF QP's.}
\label{fig1}
\end{figure}

The object of numerical studies is to diagonalize the electron-electron
interaction hamiltonian $H$ in the space of degenerate antisymmetric $N$ 
electron states of a given (lowest) LL.
Although matrix $H$ is easily block diagonalized into blocks with specified 
$M$, the exact diagonalization becomes difficult (matrix dimension over 
$10^6$) for $N>10$ and $2S>27$ ($\nu=1/3$)\cite{wojs1}.
Typical results for ten electrons at filling factors near $\nu=1/3$ are 
presented in Fig.~\ref{fig1}.
Energy $E$, plotted as a function of $L$ in the magnetic units, includes 
shift $-(Ne)^2/2R$ due to charge compensating background.
There is always one or more $L$ multiplets (marked with open circles) 
forming a low energy band separated from the continuum by a gap.
If the lowest band consists of a single $L=0$ GS (Fig.~\ref{fig1}d),
it is expected to be incompressible in the thermodynamic limit (for 
$N\rightarrow\infty$ at the same $\nu$) and an infinite 2DEG at this 
filling factor is expected to exhibit the FQHE.

The MFCF interpretation of the spectra in Fig.~\ref{fig1} is the following.
The effective magnetic monopole strength seen by CF's is\cite{jain,wojs1}
\begin{equation}
   2S^*=2S-2p(N-1),
\end{equation}
and the angular momenta of lowest CF shells (CF LL's) are $l^*_n=|S^*|+n$
\cite{chen}.
At $2S=27$, $l_0^*=9/2$ and ten CF's fill completely the lowest CF shell 
($L=0$ and $\nu^*=1$).
The excitations of the $\nu^*=1$ CF GS involve an excitation of at least
one CF to a higher CF LL, and thus (if the CF-CF interaction is weak on 
the scale of $\hbar\omega_c^*$) the $\nu^*=1$ GS is incompressible and 
so is Laughlin\cite{laughlin} $\nu=1/3$ GS of underlying electrons.
The lowest lying excited states contain a pair of QP's: a quasihole (QH) 
with $l_{\rm QH}=l^*_0=9/2$ in the lowest CF LL and a quasielectron (QE) 
with $l_{\rm QE}=l^*_1=11/2$ in the first excited one.
The allowed angular momenta of such pair are $L=1$, 2, \dots, 10.
The $L=1$ state usually has high energy and the states with $L\ge2$ form 
a well defined band with a magnetoroton minimum at a finite value of $L$.
The lowest CF states at $2S=26$ and 28 contain a single QE and a single 
QH, respectively (in the $\nu^*=1$ CF state, i.e. the $\nu=1/3$ electron
state), both with $l_{\rm QP}=5$, and the excited states will contain 
additional QE-QH pairs.
At $2S=25$ and $29$ the lowest bands correspond to a pair of QP's, and 
the values of energy within those bands define the QP-QP interaction
pseudopotential $V_{\rm QP}$.
At $2S=25$ there are two QE's each with $l_{\rm QE}=9/2$ and the allowed
angular momenta (of two identical Fermions) are $L=0$, 2, 4, 6, and 8,
while at $2S=29$ there are two QH's each with $l_{\rm QH}=11/2$ and 
$L=0$, 2, 4, 6, 8, and 10.
Finally, at $2S=24$, the lowest band contains three QE's each with 
$l_{\rm QE}=4$ in the Laughlin $\nu=1/3$ state, (in the Fermi liquid 
picture, interacting with one another through $V_{\rm QE}$\cite{sitko2}) 
and $L=1$, $3^2$, 4, 5, 6, 7, and 9.

\section{Pseudopotential Approach}

The two body interaction hamiltonian $H$ can be expressed as
\begin{equation}
   \hat{H}=\sum_{i<j} \sum_{L'} V(L')\;\hat{\cal P}_{ij}(L'),
\end{equation}
where $V(L')$ is the interaction pseudopotential\cite{haldane1} and 
$\hat{\cal P}_{ij}(L')$ projects onto the subspace with angular momentum 
of pair $ij$ equal to $L'$.
For electrons confined to a LL, $L'$ measures the average squared 
distance $d^2$\cite{wojs1},
\begin{equation}
   {\hat{d}^2\over R^2}=
   2+{S^2\over l(l+1)}\left(2-{\hat{L}'^2\over l(l+1)}\right),
\label{eqharm}
\end{equation}
and larger $L'$ corresponds to smaller separation.
Due to the confinement of electrons to one (lowest) LL, interaction 
potential $V(r)$ enters hamiltonian $H$ only through a small number 
of pseudopotential parameters $V(2l-{\cal R})$, where ${\cal R}$, 
relative pair angular momentum, is an odd integer.

\begin{figure}[t]
\epsfxsize=4in
\centerline{\epsffile{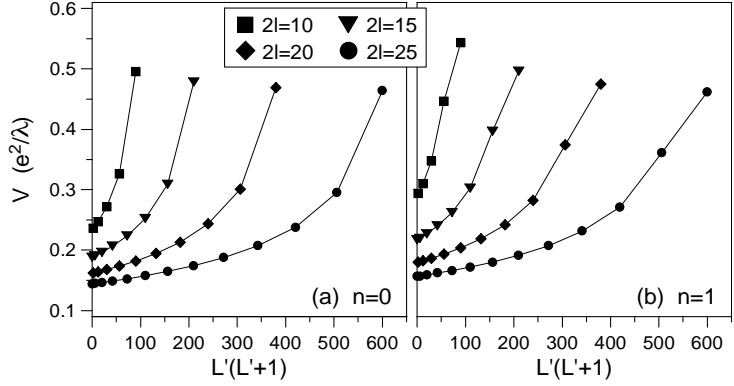}}
\caption{
   Pseudopotentials $V$ of the Coulomb interaction in the lowest (a), 
   and first excited LL (b) as a function of squared pair angular 
   momentum $L'(L'+1)$.
   Squares ($l=5$), triangles ($l=15/2$), diamonds ($l=10$), and 
   circles ($l=25/2$) mark data for different $S=l+n$.}
\label{fig2}
\end{figure}

In Fig.~\ref{fig2} we compare Coulomb pseudopotentials $V(L')$ calculated 
for a pair of electrons on the Haldane sphere each with $l=5$, 15/2, 10, 
and 25/2, in the lowest and first excited LL.
For the reason that will become clear later, $V(L')$ is plotted as 
a function of $L'(L'+1)$.
All pseudopotentials in Fig.~\ref{fig2} increase with increasing $L'$.
If $V(L')$ increased very quickly with increasing $L'$ (we define ideal 
SR repulsion as: $dV_{\rm SR}/dL'\gg0$ and $d^2V_{\rm SR}/dL'^2\gg0$), 
the low lying many body states would be the ones maximally avoiding pair 
states with largest $L'$\cite{haldane1,wojs1}.
At filling factor $\nu=1/m$ ($m$ is odd) the many body Hilbert space 
contains exactly one multiplet in which all pairs completely avoid 
states with $L'>2l-m$.
This multiplet is the $L=0$ incompressible Laughlin state\cite{laughlin} 
and it is an exact GS of $V_{\rm SR}$.

The ability of electrons in a given many body state to avoid strongly 
repulsive pair states can be conveniently described using the idea of 
fractional parentage\cite{wojs1,shalit}. 
An antisymmetric state $\left|l^N,L\alpha\right>$ of $N$ electrons each 
with angular momentum $l$ that are combined to give total angular 
momentum $L$ can be written as
\begin{equation}
   \left|l^N,L\alpha\right>=
   \sum_{L'}\sum_{L''\alpha''} 
   G_{L\alpha,L''\alpha''}(L')
   \left|l^2,L';l^{N-2},L''\alpha'';L\right>.
\label{eq3}
\end{equation}
Here, $\left|l^2,L';l^{N-2},L''\alpha'';L\right>$ denote product states 
in which $l_1=l_2=l$ are added to obtain $L'$, $l_3=l_4=\dots=l_N=l$ are 
added to obtain $L''$ (different $L''$ multiplets are distinguished by 
a label $\alpha''$), and finally $L'$ is added to $L''$ to obtain $L$.
The state $\left|l^N,L\alpha\right>$ is totally antisymmetric, and states 
$\left|l^2,L';l^{N-2},L''\alpha'';L\right>$ are antisymmetric under 
interchange of particles 1 and 2, and under interchange of any pair of 
particles 3, 4, \dots\ $N$.
The factor $G_{L\alpha,L''\alpha''}(L')$ is called the coefficient of 
fractional grandparentage (CFGP).
The two particle interaction matrix element expressed through CFGP's is
\begin{equation}
   \left<l^N,L\alpha\right|V\left|l^N,L\beta\right>
   ={N(N-1)\over2}\sum_{L'}\sum_{L''\alpha''} 
   G_{L\alpha,L''\alpha''}(L') G_{L\beta,L''\alpha''}(L') \, V(L'),
\end{equation}
and expectation value of energy is
\begin{equation}
   E_\alpha(L)={N(N-1)\over2}
   \sum_{L'} {\cal G}_{L\alpha}(L') \, V(L'),
\label{eq7}
\end{equation}
where the coefficient
\begin{equation}
   {\cal G}_{L\alpha}(L')=
   \sum_{L''\alpha''}\left|G_{L\alpha,L''\alpha''}(L')\right|^2
\label{eq52}
\end{equation}
gives the probability that pair $ij$ is in the state with $L'$.

\section{Energy Spectra of Short Range Pseudopotentials}

The very good description of actual GS's of a 2DEG at fillings $\nu=1/m$ 
by the Laughlin wavefunction (overlaps typically larger that 0.99) and 
the success of the MFCF picture at $\nu<1$ both rely on the fact that 
pseudopotential of Coulomb repulsion in the lowest LL falls into the 
same class of SR pseudopotentials as $V_{\rm SR}$.
Due to a huge difference between all parameters $V_{\rm SR}(L')$, the
corresponding many body hamiltonian has the following hidden symmetry:
the Hilbert space ${\cal H}$ contains eigensubspaces ${\cal H}_p$ of 
states with ${\cal G}(L')=0$ for $L'>2(l-p)$, i.e. with $L'<2(l-p)$.
Hence, ${\cal H}$ splits into subspaces $\tilde{\cal H}_p={\cal H}_p
\setminus{\cal H}_{p+1}$, containing states that do not have 
grandparentage from $L'>2(l-p)$, but have some grandparentage from 
$L'=2(l-p)-1$,
\begin{equation}
   {\cal H}=\tilde{\cal H}_0\oplus\tilde{\cal H}_1\oplus
            \tilde{\cal H}_2\oplus\dots
\end{equation}
The subspace $\tilde{\cal H}_p$ is not empty (some states with $L'<2(l-p)$ 
can be constructed) at filling factors $\nu\le(2p+1)^{-1}$.
Since the energy of states from each subspace $\tilde{\cal H}_p$ is 
measured on a different scale of $V(2(l-p)-1)$, the energy spectrum 
splits into bands corresponding to those subspaces.
The energy gap between the $p$th and $(p+1)$st bands is of the order of 
$V(2(l-p)-1)-V(2(l-p-1)-1)$, and hence the largest gap is that between 
the 0th band and the 1st band, the next largest is that between the 1st 
band and 2nd band, etc.

\begin{figure}[t]
\epsfxsize=4in
\centerline{\epsffile{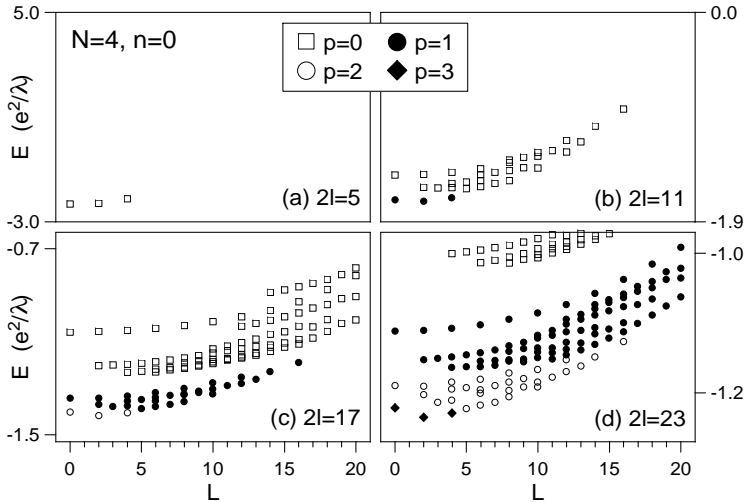}}
\caption{
   Energy spectra of four electrons in the lowest LL each with angular 
   momentum $l=5/2$ (a), $l=11/2$ (b), $l=17/2$ (c), and $l=23/2$ (d).
   Different subspaces ${\cal H}_p$ are marked with squares ($p=0$),
   full circles ($p=1$), open circles ($p=2$), and diamonds ($p=3$).}
\label{fig3}
\end{figure}

Fig.~\ref{fig3} demonstrates on the example of four electrons to what 
extent this hidden symmetry holds for the Coulomb pseudopotential in 
the lowest LL.
The subspaces ${\cal H}_p$ are identified by calculating CFGP's of 
all states.
They are not exact eigenspaces of the Coulomb interaction, but the mixing 
between different ${\cal H}_p$ is weak and the coefficients ${\cal G}(L')$ 
for $L'>2(l-p)$ (which vanish exactly in exact subspaces ${\cal H}_p$) 
are indeed much smaller in states marked with a given $p$ than in all 
other states.
For example, for $2l=11$, ${\cal G}(10)<0.003$ for states marked with 
full circled, and ${\cal G}(10)>0.1$ for all other states (squares).

Note that the set of angular momentum multiplets which form subspace 
$\tilde{\cal H}_p$ of $N$ electrons each with angular momentum $l$ is 
always the same as the set of multiplets in subspace $\tilde{\cal 
H}_{p+1}$ of $N$ electrons each with angular momentum $l+(N-1)$.
When $l$ is increased by $N-1$, an additional band appears at high 
energy, but the structure of the low energy part of the spectrum is 
completely unchanged.
For example, all three allowed multiplets for $l=5/2$ ($L=0$, 2, and 4)
form the lowest energy band for $l=11/2$, 17/2, and 23/2, where they 
span the $\tilde{\cal H}_1$, $\tilde{\cal H}_2$ and $\tilde{\cal H}_3$ 
subspace, respectively.
Similarly, the first excited band for $l=11/2$ is repeated for $l=17/2$ 
and 23/2, where it corresponds to $\tilde{\cal H}_1$ and $\tilde{\cal 
H}_2$ subspace, respectively.

Let us stress that the fact that identical sets of multiplets occur in 
subspace $\tilde{\cal H}_p$ for a given $l$ and in subspace $\tilde{\cal 
H}_{q+1}$ for $l$ replaced by $l+(N-1)$, does not depend on the form of 
interaction, and follows solely from the rules of addition of angular 
momenta of identical Fermions.
However, if the interaction pseudopotential has SR, then:
(i) $\tilde{\cal H}_p$ are interaction eigensubspaces;
(ii) energy bands corresponding to $\tilde{\cal H}_p$ with higher $p$ 
lie below those of lower $p$;
(iii) spacing between neighboring bands is governed by a difference
between appropriate pseudopotential coefficients; and
(iv) wavefunctions and structure of energy levels within each band are
insensitive to the details of interaction.
Replacing $V_{\rm SR}$ by a pseudopotential that increases more slowly
with increasing $L'$ leads to:
(v) coupling between subspaces $\tilde{\cal H}_p$;
(vi) mixing, overlap, or even order reversal of bands;
(vii) deviation of wavefunctions and the structure of energy levels 
within bands from those of the hard core repulsion (and thus their
dependence on details of the interaction pseudopotential).
The numerical calculations for the Coulomb pseudopotential in the lowest 
LL show (to a large extent) all SR properties (i)--(iv), and virtually 
no effects (v)--(vii), characteristic of 'non SR' pseudopotentials.

The reoccurrence of $L$ multiplets forming the low energy band when 
$l$ is replaced by $l\pm(N-1)$ has the following crucial implication.
In the lowest LL, the lowest energy ($p$th) band of the $N$ electron 
spectrum at the monopole strength $2S$ contains $L$ multiplets which 
are all the allowed $N$ electron multiplets at $2S-2p(N-1)$.
But $2S-2p(N-1)$ is just $2S^*$, the effective monopole strength of CF's!
The MFCS transformation which binds $2p$ fluxes (vortices) to each 
electron selects the same $L$ multiplets from the entire spectrum as 
does the introduction of a hard core, which forbids a pair of electrons 
to be in a state with $L'>2(l-p)$.

\section{Definition of Short Range Pseudopotential}

A useful operator identity relates total ($L$) and pair ($\hat{L}_{ij}$) 
angular momenta\cite{wojs1}
\begin{equation}
   \sum_{i<j} \hat{L}_{ij}^2 = \hat{L}^2 + N(N-2)\;\hat{l}^2.
\label{eqthr}
\end{equation}
It implies that interaction given by a pseudopotential $V(L')$ that is 
linear in $\hat{L}'^2$ (e.g. the harmonic repulsion within each LL; see
Eq.~(\ref{eqharm})) is degenerate within each $L$ subspace and its energy 
is a linear function of $L(L+1)$.
The many body GS has the lowest available $L$ and is usually degenerate,
while the state with maximum $L$ has the largest energy.
Note that this result is opposite to the Hund rule valid for spherical
harmonics, due to the opposite behavior of $V(L')$ for the FQH ($n=0$ 
and $l=S$) and atomic ($S=0$ and $l=n$) systems.

Deviations of $V(L')$ from a linear function of $L'(L'+1)$ lead to 
the level repulsion within each $L$ subspace, and the GS is no longer 
necessarily the state with minimum $L$.
Rather, it is the state at a low $L$ whose multiplicity $N_L$ (number of 
different $L$ multiplets) is large.
It interesting to observe that the $L$ subspaces with relatively high 
$N_L$ coincide with the MFCF prediction.
In particular, for a given $N$, they reoccur at the same $L$'s when 
$l$ is replaced by $l\pm(N-1)$, and the set of allowed $L$'s at 
a given $l$ is always a subset of the set at $l+(N-1)$.

As we said earlier, if $V(L')$ has SR, the lowest energy states within 
each $L$ subspace are those maximally avoiding large $L'$, and the 
lowest band (separated from higher states by a gap) contains states 
in which a number of largest values of $L'$ is avoided altogether.
This property is valid for all pseudopotentials which increase more
quickly than linearly as a function of $L'(L'+1)$.
For $V_\beta(L')=[L'(L'+1)]^\beta$, exponent $\beta>1$ defines the 
class of SR pseudopotentials, to which the MFCF picture can be applied.
Within this class, the structure of low lying energy spectrum and the
corresponding wavefunctions very weakly depend on $\beta$ and converge to 
those of $V_{\rm SR}$ for $\beta\rightarrow\infty$.

The extension of the SR definition to $V(L')$ that are not strictly in 
the form of $V_\beta(L')$ is straightforward.
If $V(L')>V(2l-m)$ for $L'>2l-m$ and $V(L')<V(2l-m)$ for $L'<2l-m$ and
$V(L')$ increases more quickly than linearly as a function of $L'(L'+1)$ 
in the vicinity of $L'=2l-m$, then pseudopotential $V(L')$ behaves like 
a SR one at filling factors near $\nu=1/m$.

\section{Application to Various Pseudopotentials}

It follows from Fig.~\ref{fig2}a that the Coulomb pseudopotential in 
the lowest LL satisfies the SR condition in the entire range of $L'$;
this is what validates the MFCF picture for filling factors $\nu\le1$.
It also explains the formation of incompressible states of charged
magneto-excitons ($X^-$) formed in the electron-hole plasma\cite{wojs2}.
However, in a higher, $n$th LL this is only true for $L'<2(l-n)-1$
(see Fig.~\ref{fig2}b for $n=1$) and the MFCF picture is valid only for 
$\nu_n$ (filling factor in the $n$th LL) around and below $(2n+3)^{-1}$.
Indeed, the MFCF features in the ten electron energy spectra around 
$\nu=1/3$ (in Fig.~\ref{fig1}) are absent for the same fillings of the 
$n=1$ LL \cite{wojs1}.

\begin{figure}[t]
\epsfxsize=4in
\centerline{\epsffile{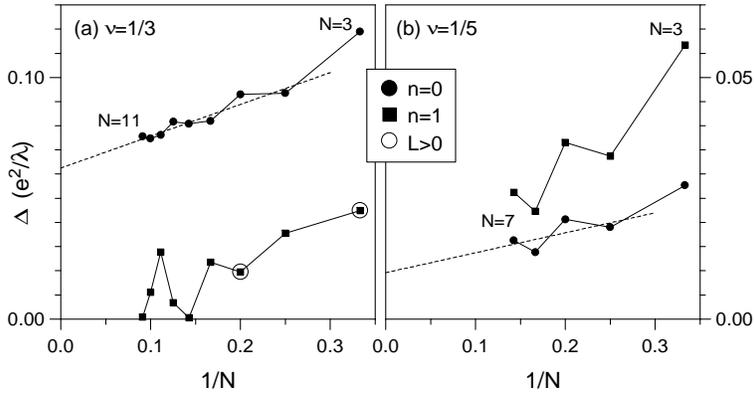}}
\caption{
   Excitation gap $\Delta$ as a function of inverse electron number 
   $1/N$ for filling factors $\nu=1/3$ (a) and 1/5 (b) in the $n=0$ 
   (dots) and $n=1$ (squares) LL's. 
   Open circles mark degenerate ground states ($L>0$).}
\label{fig4}
\end{figure}

One consequence of this is that the MFCF picture or Laughlin like 
wavefunction cannot be used to describe the reported\cite{willet} 
incompressible state at $\nu=2+1/3=7/3$ ($\nu_1=1/3$).
The correlations in the $\nu=7/3$ GS are different than at $\nu=1/3$;
the origin of (apparent) incompressibility cannot be attributed to the 
formation of a Laughlin like $\nu_1=1/3$ state (in which pair states 
with smallest average separation $d^2$ are avoided) on top of the 
$\nu=2$ state and connection between the excitation gap and the 
pseudopotential parameters is different.
This is clearly visible in the dependence of the excitation gap $\Delta$
on the electron number $N$, plotted in Fig.~\ref{fig4} for $\nu=1/3$ and 
1/5 fillings of the lowest and first excited LL.
The gaps for $\nu=1/5$ behave very similarly as a function of $N$ in 
both LL's, while it is not even possible to make a conclusive statement
about degeneracy or incompressibility of the $\nu=7/3$ state based on
our data for up to eleven electrons.

The SR criterion can be applied to the QP pseudopotentials to understand
why QP's do not form incompressible states at all Laughlin filling factors 
$\nu_{\rm QP}=1/m$ in the hierarchy picture\cite{haldane2,sitko1} of FQH 
states.
Lines in Fig.~\ref{fig1}b and f mark $V_{\rm QE}$ and $V_{\rm QH}$ for 
the Laughlin $\nu=1/3$ state of ten electrons.
Clearly, the incompressible states with a large gap will be formed by
QH's at $\nu_{\rm QH}=1/3$ and by QE's at $\nu_{\rm QE}=1$, explaining
strong FQHE of the underlying electron system at Jain $\nu=2/7$ and 2/5
fractions, respectively.
On the other hand, there is no FQHE at $\nu_{\rm QH}=1/5$ ($\nu=4/13$) 
or $\nu_{\rm QE}=1/3$ ($\nu=4/11$), and the gap above possibly 
incompressible $\nu_{\rm QH}=1/7$ ($\nu=6/19$) and $\nu_{\rm QE}=1/5$ 
($\nu=6/17$) states should be very small, which agrees very well with
exact few electron calculations.
We believe that taking into account the behavior of involved QP 
pseudopotentials on all levels of hierarchy should explain all observed 
odd denominator FQH fillings and allow prediction of their relative 
stability (without using trial wavefunctions involving multiple LL's and 
projections onto the lowest LL needed in the Jain\cite{jain} picture).

\section{Conclusion}

Using the pseudopotential formalism, we have described the FQH states 
in terms of the ability of electrons to avoid strongly repulsive pair 
states.
We have defined the class of SR pseudopotentials leading to the formation 
of incompressible FQH states.
We argue that the MFCF picture is justified for the SR interactions 
and fails for others.
The pseudopotentials of the Coulomb interaction in excited LL's and 
of Laughlin QP's in the $\nu=1/3$ state are shown to belong to the 
SR class only at certain filling factors.

\section{Acknowledgment}
This work has been supported in part by the Materials Research Program 
of Basic Energy Sciences, US Department of Energy.
A.W. thanks Witold Bardyszewski (Warsaw University) for help with
improving the numerical codes.

\end{document}